\documentstyle[11pt,aaspp4,flushrt]{article}
\input psfig
\begin{document}
\title{Comment on ``Dynamic Screening...'' by Opher and Opher} 
\author{Andrei Gruzinov}
\affil{Institute for Advanced Study, School of Natural Sciences, Princeton, NJ 08540}

\begin{abstract}
I have shown that the enhancement of fusion rates in stars due to electrostatic screening does not depend on the Gamow energy of the fusing nuclei (in the classical weak screening limit). Opher and Opher have recently published an ApJ paper claiming that my proof is wrong. I show here that their arguments are incorrect.
\end{abstract}

\section{Introduction}
The rate of thermonuclear reactions in stars is increased by the Debye-Huckle screening because the screening reduces the Coulomb repulsion. In the weak screening limit, which is the first classical approximation, the reaction rate enhancement factor $w$ was calculated by Salpeter (1954): $w=1+\Lambda$, where $\Lambda\equiv Z_1Z_2e^2/TR_D$, $Z_{1,2}$ are the charges of the fusing nuclei, $T$ is the temperature, $R_D$ is the Debye radius. 

Thirty-four years later Carraro, Schafer, \& Koonin (1988) made an interesting suggestion. They noticed that the Gamow energy of the reacting nuclei is high, i.e., only rare fast-moving nuclei have a noticeable chance to fuse. Fast ions will induce a smaller electrostatic response in the plasma than assumed by Salpeter. The authors proposed that Salpeter's weak screening formula is not strictly valid because $\Lambda$ actually depends on the Gamow energy of the reaction ($\Lambda$ decreases with the increasing Gamow energy). The phenomenon was termed dynamic screening. Stellar evolution people started to use the modified fusion rates in their numerical codes.

I have recently explained that this effect is actually absent (Gruzinov, 1998). In the classical weak screening limit, $w$ does not depend on the Gamow energy of the fusing nuclei, and is given by the Salpeter formula. My explanation boils down to the following. The weak screening formula of Salpeter can be derived in the framework of classical statistical mechanics (e.g. DeWitt, Graboske, \& Cooper, 1973). In classical statistical mechanics, fast nuclei are just as screened as slow nuclei, because the Gibbs distribution factorizes into kinetic and configuration parts. Therefore, the Gamow energy cannot enter the expression for the screening enhancement factor. A physical mechanism responsible for this somewhat paradoxical velocity-independence of the screening was identified.

Now I have to return to this problem because Opher and Opher have published an ApJ paper claiming that my proof is wrong. I will explain that their arguments are incorrect.

\section{Opher \& Opher (1999)}
(i) I quote Opher \& Opher (1999) ``The exact Gibbs distribution takes into account dynamic corrections''. I quote Landau \& Lifshitz (1980) ``...the probabilities for momenta and coordinates are independent, in the sense that any particular values of the momenta do not influence the probabilities of the various values of the coordinates, and vice versa.''

(ii) In my paper I calculate the thermal electric field using the classical ($\hbar =0$) theory. Opher \& Opher (1999) propose to use $\hbar =1$. I know that $\hbar >0$, and I have a rough idea of how to calculate quantum corrections to the fusion rates' enhancement factor (Gruzinov \& Bahcall, 1998). But the point of my paper was to show that the dynamic screening effect is in fact absent. Since the paper of Carraro, Schafer, \& Koonin (1988) introduces the dynamic screening in a purely classical way, I have used the classical theory to show that the effect is spurious.

\section{Conclusion}
Today Salpeter's weak screening formula is the most reliable approximation for fusion rates in the Sun (Gruzinov \& Bahcall, 1998, Adelberger et al., 1998). The accuracy of the Salpeter's formula is not worse than few percent (the calculated fusion rates deviate from the real fusion rates by no more than a few percent for all relevant solar nuclear reactions).

\acknowledgements This work was supported by NSF PHY-9513835.

\end{document}